\documentclass[twocolumn,showpacs,preprintnumbers,amsmath,amssymb]{revtex4}

\usepackage{epsfig}
\usepackage{graphicx}
\usepackage{dcolumn}
\usepackage{bm}
\usepackage{epstopdf}
\begin{document}

\title{\bf Transport and Helfand moments in the Lennard-Jones fluid. 
I. Shear viscosity}

\author{S. Viscardy, J. Servantie, and P. Gaspard\\
{\em Center for Nonlinear Phenomena and Complex Systems,}\\
{\em Universit\'e Libre de Bruxelles,}\\
{\em Campus Plaine, Code Postal 231, B-1050 Brussels, Belgium}\\}

\begin{abstract}
We propose a new method, the \textit{Helfand-moment method}, to 
compute the shear viscosity
by equilibrium molecular dynamics in periodic systems. In this 
method, the shear viscosity is written as an Einstein-like relation
in terms of the variance of the so-called \textit{Helfand moment}. 
This quantity,
is modified in order to satisfy systems with periodic boundary 
conditions usually considered in molecular dynamics.
We calculate the shear viscosity in the Lennard-Jones fluid near the 
triple point thanks to this new technique.
We show that the results of the Helfand-moment method are in excellent 
agreement with the results of the standard Green-Kubo method.
\end{abstract}


\maketitle

\section{Introduction}

Since Maxwell's first paper \cite{maxwell-1860,Viscardy-HS} on the 
kinetic theory of gases,
shear viscosity as well as the other 
transport properties are known to find their origin
in the 
microscopic motion of atoms and molecules composing matter.
However, 
it is only since the fifties that exact formulas are known 
to
calculate the transport coefficients in terms of the microscopic 
dynamics.
These so-called Green-Kubo formulas give each transport 
coefficient as the time
integral of the autocorrelation function of 
some specific microscopic flux
associated with the transport property 
of interest \cite{green51,green60,kubo57,mori58} .
Today, the 
Green-Kubo technique allows us to calculate numerically
the transport 
coefficients by simulating the molecular dynamics of
systems with a 
finite number of particles and periodic boundary conditions.

On the 
other hand, Einstein classic work on Brownian motion showed that 
transport properties
such as diffusion can also be understood in 
terms of random walks.  It was Helfand \cite{helf}
who identified in 
1960 the fluctuating quantities which, by their random walk, are 
associated
with each transport coefficients.  These fluctuating 
quantities are the so-called Helfand moments
and are the 
centroids of the conserved quantity which is transported.
In 
principle, each transport coefficient can thus be obtained from the 
linear increase of the
statistical variance of the corresponding 
Helfand moment by the so-called generalized Einstein relations. 
Nevertheless, it is not yet known today how these Helfand moments 
should be defined in molecular dynamics with periodic boundary 
conditions, which limits their use in numerical simulations.

The 
purpose of the present paper is to derive an analytical expression of 
the Helfand moment
associated with viscosity for molecular dynamics 
with periodic boundary conditions,
and to apply the Helfand-moment 
method to the calculation of shear viscosity
in the Lennard-Jones 
fluid near the triple point.

Already in the first numerical 
calculation of viscosity in 1970 \cite{alder},
the algorithm of Alder 
\textit{et al.} was based on generalized Einstein relations 
derived 
from the Green-Kubo formulas.  The application of the pure Green-Kubo 
technique by equilibrium molecular dynamics to the Lennard-Jones 
fluid has
been performed a short time after by Levesque \textit{et al} 
\cite{LVK-73} and later by Schoen and Hoheisel 
\cite{schoen-hoheisel85}.
However, until the middle of the eighties and the work of Schoen and 
Hoheisel \cite{schoen-hoheisel85}, as well as the one of Erpenpeck 
using the  Monte-Carlo Metropolis method \cite{erpenbeck-88}, 
nonequilibrium molecular dynamics was predominantly used for the 
computation
of shear viscosity 
\cite{lees-edwards72,ashurst-hoover73,HEHLAM80,evans81,trozzi-ciccotti84}.

At the end of the eighties and the beginning of the nineties, the 
generalized Einstein relations
started to be used for calculating the 
transport coefficients. 
An alternative equilibrium molecular 
dynamics method 
has been proposed in which the variance of the time 
integral of the microscopic flux 
is calculated \cite{allen93,haile}.
Actually it is the analog of the method of Alder \textit{et al.}'s 
\cite{alder} for soft sphere potential systems. Recently, this 
technique has been applied by Meier \textit{et al.} \cite{MLK-04}, 
and by Hess and Evans \cite{HE01}, the latter having rather 
considered an equilibrium ensemble of time averages of the flux. 
In 
this context, two important points were discussed.
The first concerns 
the so-called \textit{McQuarrie expression}. In his book 
\cite{mcquarrie}, McQuarrie presented Helfand's formula for the shear 
viscosity, but with a slightly different
form. This difference implied a simplification of the expression, 
apparently giving an important advantage compared to the original 
formula \cite{CD91,CCE93,ABM94,allen94}.
The second point concerned the validity of the generalized Einstein 
relations in periodic systems \cite{haile,allen93,ABM94,erpenbeck}. 
Arguing that the periodic boundary conditions imply that the variance 
of the original expression of the \textit{Helfand moment} is bounded 
in time, the generalized Einstein relations were considered to be 
impractical in periodic systems. In this paper, we will show that a 
generalized Einstein relation is available for viscosity after the 
addition of two terms to the original Helfand moment to take into 
account the periodicity of the system. The Helfand-moment method 
presents the important advantage to define shear viscosity as a 
non-negative quantity, satisfying the positivity of the entropy 
production. We have previously applied such a method to a system of 
two hard disks with periodic boundary conditions 
\cite{viscardy-gaspard1}. In the present paper, we calculate the 
shear viscosity in a Lennard-Jones fluid near the triple point. We 
compare the results obtained by the Helfand-moment method with our 
own Green-Kubo values and those found in the literature.

In addition, the Helfand-moment method plays an important role in the 
\textit{escape-rate formalism}. This formalism establishes direct 
relationships between the characteristic quantities of the 
microscopic chaos (Lyapunov exponents and fractal dimensions) and the 
transport coefficients 
\cite{dorf-gasp,gasp-dorf,gasp-book,dorf-book}. A few years ago, such 
a relation has been studied for the viscosity in the two-hard-disk 
model \cite{viscardy-gaspard2}. Furthermore with the use of the 
Helfand moment, it should be possible to construct at the microscopic 
level the hydrodynamic modes, which are the solutions of the 
Navier-Stokes equations. This approach called the hydrodynamic-mode 
method has been successfully applied for diffusion \cite{gaspard96}.

The paper is organized as follows. In Section \ref{EH}, we outline 
the theoretical background of the generalized Einstein formula.
Section \ref{PBC} is devoted to the presentation of our 
Helfand-moment method used for the calculation of the shear viscosity 
in this paper. In Section \ref{Discussion}, we discuss the so-called 
McQuarrie expression and the validity of the generalized Einstein 
formula for periodic systems.
The results of the molecular dynamics simulations are given in 
Section \ref{Num}. The comparison with our Green-Kubo results and 
previous researches are done. Finally, conclusions are drawn in 
Section \ref{Conclusions}.

%%%%%%%%%%%%%%%%%%%%%%%%%%%%%%%%%%%%%%%%%%%%%%%%%%%%%%%%%%%%%%%%%%%%%%%%%%%%%%%%

\section{Einstein-Helfand formula}
\label{EH}

One century ago, Einstein theoretically established a relationship 
between the diffusion coefficient of Brownian motion and the random 
walk of the Brownian particle due to its collisions with the 
molecules of the surrounding fluid \cite{einstein}:
\begin{equation}
D = \lim_{t \to \infty} \frac{\left \langle \left \lbrack x(t) - x(0) 
\right \rbrack ^2 \right \rangle}{2 t}\, ,
\label{einstein.diffusion}
\end{equation}
where $x$ is the position of the colloidal particle and $t$ the time. 
Thereafter, different studies led to the well-known Green-Kubo 
formula for
the shear viscosity $\eta$ obtained by Green \cite{green51,green60}, 
Kubo \cite{kubo57}  and Mori \cite{mori58}:
\begin{equation}
\eta = \lim_{N,V\to\infty} \frac{1}{k_{\rm B} T V} \int_{0}^{\infty} 
dt \left \langle J^{(\eta)}(t) J^{(\eta)}(0)\right \rangle  ,
\label{transp-coeff-GK}
\end{equation}
where $J^{(\eta)}$ is the microscopic flux associated with the shear 
viscosity $\eta$.
The viscosity coefficient is obtained in the 
thermodynamic limit where $N,V\to\infty$ while the particle density 
$n=N/V$ remains constant.  In the following, this condition is always 
assumed together with the limit $N,V\to\infty$.
The simplicity of Eq. (\ref{einstein.diffusion}) obtained by Einstein 
presents a particular interest. The extension of such a relation to 
the other transport coefficients could be useful. In the sixties, 
Helfand proposed quantities associated with the different transport 
processes in order to establish Einstein-like relations for the 
transport coefficients \cite{helf}. In particular, we have for shear
viscosity that
\begin{equation}
\eta=\lim_{N,V,t \to \infty}\frac{1}{2k_{\rm B} T Vt} \left \langle 
\left[G^{(\eta)}
(t)-G^{(\eta)} (0) \right]^{2}\right \rangle .
\label{Einstein.shear.viscosity}
\end{equation}
It can be shown \cite{helf} that the Einstein-Helfand relation is 
equivalent 
to the Green-Kubo formula (\ref{transp-coeff-GK}) by 
defining
the Helfand moment as:
\begin{equation}
G^{(\eta)}(t) = \sum_{a=1}^{N} p_{xa}(t) y_a (t) \, ,
\label{helfand-moment-shear-viscosity}
\end{equation}
if the corresponding microscopic flux is the time derivative of the 
Helfand moment:
\begin{equation}
J^{(\eta)}(t)=\frac{d}{dt} G^{(\eta)}(t) \, .
\label{derivativeG}
\end{equation} 
We notice that the limit $t\to\infty$ should be 
related to the thermodynamic limit $N,V\to\infty$.
Indeed, for a 
fluid of particles confined in a finite box, the quantity 
(\ref{helfand-moment-shear-viscosity}) is bounded so that the 
coefficient (\ref{Einstein.shear.viscosity}) would vanish if
the 
limit $t\to\infty$ was taken before the thermodynamic limit 
$N,V\to\infty$.  Therefore, the number $N$ of particles and the 
volume $V$ should be large enough in order that the variance of the 
Helfand moment displays a linear increase over a sufficiently long 
time interval $t$, allowing the coefficient $\eta$ to be well 
defined.  The larger the system, the longer the time interval.  It is 
in this sense that the limit $N,V,t \to \infty$ should be considered 
in Eq. (\ref{Einstein.shear.viscosity}).

Another remark is that, 
compared with the Green-Kubo formula (\ref{transp-coeff-GK}), Eq. 
(\ref{Einstein.shear.viscosity})
presents the advantage to define the shear viscosity as a positive 
quantity, satisfying the conditions of non-negative entropy 
production.

%%%%%%%%%%%%%%%%%%%%%%%%%%%%%%%%%%%%%%%%%%%%%%%%%%%%%%%%%%%%%%%%%%%%%%%%%%%%%%%%

\section{Helfand moment in periodic systems}
\label{PBC}

Often, the molecular dynamics is simulated with periodic boundary 
conditions.
In this case, the particles exiting at one boundary are 
reinjected at the opposite boundary.
Due to the periodicity of the system, particles in the simulation box may interact with image particles
as well as the particles inside the original unit 
cell. As a consequence, the images of
particle $b$ may contribute to the force ${\bf F}_{ab}$ applied by 
the particle $b$ on the particle $a$:
\begin{equation}
{\bf F}_{ab} = - \sum_{{\pmb \beta}^{(a,b)}}\frac{\partial u({\bf 
r}_{ab})}{\partial {\bf r}_{ab}}
\label{periodic-force}
\end{equation}
with
\begin{equation}
{\bf r}_{ab} = {\bf r}_a - {\bf r}_b - L\;  {\pmb\beta}^{(a,b)} 
\label{MIC}
\end{equation}
where $L$ the length of the simulation box and ${\pmb\beta}^{(a,b)}$ 
determines the cell translation vector \cite{haile}. The range of the 
interaction potential must be smaller than $L/2$ to guarantee that 
the particle $a$ interacts only with one of the images of $b$ in Eq. 
(\ref{periodic-force}). The interacting pair is found by the 
minimum-image convention, $\Vert{\bf r}_a - {\bf r}_b - L\, 
{\pmb\beta}^{(a,b)}\Vert < L/2$. Here, we define the quantity
\begin{equation}
{\bf L}_{b|a}(t) = L \; {\pmb\beta}^{(a,b)} (t)
\label{definition-Lab}
\end{equation}
which is the vector to be added to ${\bf r}_b$ in order to satisfy 
the minimum-image convention. 

For a dynamics which is periodic in 
the box, the positions should jump to satisfy the minimum-image 
convention.  As a consequence, the positions and momenta
used to calculate the viscosity by the Green-Kubo method actually 
obey the modified Newtonian equations
\begin{eqnarray}
\frac{d{\bf r}_a}{dt} &=& \frac{{\bf p}_a}{m} +
\sum_{s} \Delta{\bf r}_{a}^{(s)} \; \delta(t-t_{s})\; , \nonumber \\
\frac{d{\bf p}_{a}}{dt} &=& \sum_{b (\neq a)} {\bf F}({\bf r}_{ab})  \; ,
\label{Newton}
\end{eqnarray}
where $\Delta{\bf r}_{a}^{(s)}$ is the jump of the particle $a$ at 
time $t_{s}$ with $\Vert \Delta {\bf r}_{a}^{(s)}  \Vert = L$.
We notice that the modified Newtonian equations (\ref{Newton}) 
conserve energy, total momentum and preserve phase-space volumes
(Liouville's theorem). 

Moreover, we see that the periodic boundary 
conditions imply that the Helfand moment of Eq. 
(\ref{helfand-moment-shear-viscosity}) is bounded and cannot be 
differentiated near the times $t_s$ of the jumps. In order to have a 
well-defined quantity, one should remove the discontinuities at the 
jumps, so that the Helfand moment can grow without bound. In order to do that, we add a term $I(t)$ to the original Helfand moment 
(\ref{helfand-moment-shear-viscosity}) to get
\begin{equation}
G^{(\eta)}(t) = \sum_{a}^{} p_{ax}(t) y_{a}(t) + I(t) \; .
\label{Helfand+I}
\end{equation}
According to Eq. (\ref{derivativeG}), the time derivative of the 
Helfand moment must be the microscopic flux
\begin{equation}
J^{(\eta)}(t) = \sum_{a}^{} \frac{p_{ax} p_{ay}}{m} + 
\frac{1}{2}\sum_{a \ne b} F_x({\bf r}_{ab}) \; y_{ab} \, ,
\label{microflux}
\end{equation}
defined with the position $y_{ab}$ of the 
minimum-image convention.
In order to satisfy the equality (\ref{derivativeG}) in periodic 
systems, we show in Appendix \ref{appendixA} that the term $I(t)$ 
must be given by
\begin{eqnarray}
I(t)  &=&  - \sum_{a}^{} \sum_{s}^{} p_{ax}^{(s)} \Delta y_{a}^{(s)} 
\theta (t-t_s) \nonumber \\
&-& \frac{1}{2} \sum_{a \ne b}^{} \int_{0}^{t} d \tau \; F_x ({\bf 
r}_{ab}) \; L_{b|a y}  \; .
\label{added-terms}
\end{eqnarray}
where both ${\bf r}_{ab}$ and $L_{b|a y}$ depend on 
the time $\tau$ in the integral of the last term.
We then obtain our 
general expression for the Helfand moment in systems with periodic 
boundary conditions:
\begin{eqnarray}
G^{(\eta)}(t) &=& \sum_{a=1}^N p_{ax}(t) \; y_{a}(t) \nonumber \\
&-& \sum_{a}^{} \sum_s p_{ax}^{(s)} \; \Delta y_{a}^{(s)} \; 
\theta(t-t_s) \nonumber \\
&-& \frac{1}{2} \sum_{a \ne b}^{} \int_{0}^{t} d \tau \; F_x ({\bf 
r}_{ab}) \; L_{b|a y}
\label{helfand-torus}
\end{eqnarray}
where $p_{ai}^{(s)}=p_{ai}(t_s)$ is the momentum at the time of the 
jump $t_s$ and $\theta(t-t_s)$ the Heaviside step function defined as
\begin{eqnarray}
\theta(t-t_s) = \left \lbrace
\begin{array}{cc}
  1 \;  \quad {\rm for} \ \ t>t_s \; ,\\
  0 \;  \quad {\rm for} \ \ t<t_s \; .\\
\end{array}
\right .
\end{eqnarray}
We notice that the quantity $L_{b|a y}$ has discontinuous jumps in 
order to satisfy the minimum-image convention. Let us point out that 
$L_{b|a y}$ changes when the force $F_x ({\bf r}_{ab})$ vanishes, so 
that the last term varies continuously in time and does not present 
any jump. We notice thatthe last two terms of Eq. 
(\ref{helfand-torus}) involves
  the particles near the boundaries of the box. The second term is due 
to the jumps of the particles to or from the neighboring boxes. The
  third term concerns the pairs of particles interacting between 
neighboring cells. The Helfand moment of Eq. (\ref{helfand-torus}) 
can be used
  to obtain the shear viscosity coefficient for  systems with periodic 
boundaries thanks to the Einstein-like relation 
(\ref{Einstein.shear.viscosity}).

%%%%%%%%%%%%%%%%%%%%%%%%%%%%%%%%%%%%%%%%%%%%%%%%%%%%%%%%%%%%%%%%%%%%%%%%%%%%%%%%

\section{Discussion}
\label{Discussion}

Since the beginning of the nineties, some confusions have been 
propagated in the literature concerning the use of the mean-squared 
displacement equation for shear viscosity. First, it concerns the 
so-called McQuarrie expression. On the other hand, several works have 
been done
which have prematurely concluded that the mean-square displacement 
equation for shear viscosity is inapplicable for systems with 
periodic boundary conditions. These confusions and criticisms are 
reported in particular by Erpenbeck in Ref. \cite{erpenbeck}.
Since these questions are central in this paper, this section is 
devoted to such problems in order to avoid any misconception.

\subsection{McQuarrie expression for shear viscosity}

In his well-known and remarkable book \textit{Statistical Mechanics}, 
McQuarrie  \cite{mcquarrie} reported the work achieved by Helfand 
\cite{helf}.
The derivation he proposed is quite different but he obtained the 
same intermediate relation as Helfand, that is
\cite{helfand-mcquarrie-note}
\begin{equation}
\eta = \lim_{N,V,t \to \infty}\frac{1}{2k_{\rm B} T Vt} \left \langle 
\sum_{a,b = 1}^{N} \left \lbrack x_a(t) - x_b(0) \right \rbrack^2 
p_{ay}(t) p_{by}(0) \right \rangle .
\label{Helf-McQuarrie-comparison}
\end{equation}
Thereafter in his book, McQuarrie let as an exercise the derivation 
from Eq. (\ref{Helf-McQuarrie-comparison}) of the final expression 
which is printed in Ref. \cite{mcquarrie} as follows
\begin{equation}
\eta_{\rm MQ}=\lim_{N,V,t \to \infty}\frac{1}{2k_{\rm B} T Vt} \left 
\langle  \sum_{a=1}^{N} \left[ x_{a}(t) p_{ay}(t) -  x_{a}(0) 
p_{ay}(0) \right]^{2}\right \rangle ,
\label{mcquarrie-viscosity}
\end{equation}
while Helfand obtained
\begin{equation}
\eta_{\rm H}=\lim_{N,V,t \to \infty}\frac{1}{2k_{\rm B} T Vt} \left 
\langle  \left \lbrack \sum_{a=1}^{N}  x_{a}(t) p_{ay}(t) -  x_{a}(0) 
p_{ay}(0) \right]^{2} \right \rangle .
\label{helfand-viscosity}
\end{equation}
The difference between both expressions is in the position of the sum 
over particles, and it seems that such a difference is due to
a typing error. Nevertheless, the McQuarrie expression 
(\ref{mcquarrie-viscosity}) at first sight presents a certain 
advantage compared to
Helfand's one (\ref{helfand-viscosity}). Indeed, the sum over the 
particles can come out of the average. Consequently, one would obtain 
a sum of
averages no longer depending on the different particles. If Eq. 
(\ref{mcquarrie-viscosity}) would hold, Eq. 
(\ref{mcquarrie-viscosity}) could be rewritten as
\begin{equation}
\eta_{\rm MQ}=\lim_{N,V,t \to \infty}\frac{N}{2k_{\rm B} T Vt} \left 
\langle \left[ x_{1}(t) p_{1y}(t) -  x_{1}(0) p_{1y}(0) \right]^{2}
\right \rangle .
\label{mcquarrie-viscosity-2}
\end{equation}
In other words, the McQuarrie relation seems to present the advantage 
that shear viscosity would be evaluated through a 
\textit{single-particle}
expression whereas Helfand expressed the viscosity by a 
\textit{collective} approach.

The first time that Eq. (\ref{mcquarrie-viscosity})  has been 
considered was in the work of Chialvo and Debenedetti \cite{CD91}. 
Without
giving a theoretical proof of the validity of the last equation or 
the equivalence with Eq. (\ref{helfand-viscosity}), they provided a 
numerical comparison between both methods and concluded that the 
difference between $\eta_{\rm H}$ and $\eta_{\rm MQ}$ is small.
Later, Chialvo, Cummings and Evans \cite{CCE93} also compared the 
McQuarrie and Helfand expressions and speculated an equivalencefor 
theoretical reasons. Thereafter, Allen, Brown and Masters 
\cite{ABM94} showed by comparison with their Green-Kubo results that 
the numerical calculations for shear viscosity obtained by Chialvo 
and Debenedetti \cite{CD91} were incorrect. Moreover, Allen devoted a 
comment in Ref. \cite{CCE93}, and concluded that the McQuarrie 
expression is not valid and is not able to give shear viscosity 
\cite{allen94}.
  This conclusion was confirmed later by Erpenbeck \cite{erpenbeck}, 
which settled the question. As aforementioned, this is not really 
surprising since Eq. (\ref{mcquarrie-viscosity}) seems quite clearly 
to be the result of a typing error.  This discussion emphasizes the 
fact that viscosity is a collective transport property, implying the 
intervention of all the particles.

\subsection{Periodic systems and Helfand-moment method}

The other point which was questioned is whether a mean-squared 
displacement equation for shear viscosity is useful for systems 
submitted to periodic boundary conditions 
\cite{allen93,ABM94,erpenbeck}.  First, it was pointed out that the 
well known Alder \textit{et al.} method initially developed for 
hard-ball systems \cite{alder} is not based on the Helfand 
expressions, but instead on the mean-square displacement of the time 
integral of the microscopic flux \cite{erpenbeck}.

The main doubt on the use of Helfand moments in periodic systems 
comes from the fact that the original expression 
(\ref{helfand-moment-shear-viscosity}) is bounded and would lead to a 
vanishing shear viscosity in the long-time limit.  By this argument, 
Allen concluded that
the \textit{only} correct way to handle $G_{xy}(t)$ is to write it as 
$\int_{0}^{t} \dot{G}_{xy}(\tau) \; d \tau$,
and express $ \dot{G}_{xy}$ in pairwise, minimum-image form \cite{allen93}.
In other words, the Alder \textit{et al.} method would be the only 
valid method for studying viscosity,
that is, through a method intermediate between the Helfand and 
Green-Kubo methods. Let us mention
that this opinion was recently followed by Hess, Kr\"{o}ger and Evans 
\cite{HE01,HKE03} as well as by Meier, Laesecke and Kabelac 
\cite{MLK-04,MLK-05} having considered systems with soft-potential 
interactions. However, this does not preclude the possibility to 
modify the original expression (\ref{helfand-moment-shear-viscosity}) 
of the Helfand moment in order to recover the microscopic flux 
(\ref{microflux}).  This is precisely what we have done here above 
with our Helfand-moment method by adding the following two terms
\begin{eqnarray}
&-& \sum_{a=1}^N \sum_s p_{ax}^{(s)} \; \Delta y_{a}^{(s)} \; 
\theta(t-t_s)  \nonumber \\
&-& \frac{1}{2} \sum_{a \ne b}^{} \int_{0}^{t} d \tau \; F_x ({\bf 
r}_{ab}) \; L_{b|a y}
\end{eqnarray}
to the original one.  Albeit the first original term is bounded in 
time, the two new terms increase without bound in time because of the 
jumps
and the interactions between the particle and the image particles 
(due to the minimum-image convention). Therefore, they can contribute 
to the linear growth in time of the variance of the Helfand moment. 
The Helfand-moment method we propose here is completely equivalent to 
the Green-Kubo formula and presents the advantage to express the 
transport coefficients by Einstein-like relations, directly showing 
their positivity.

%%%%%%%%%%%%%%%%%%%%%%%%%%%%%%%%%%%%%%%%%%%%%%%%%%%%%%%%%%%%%%%%%%%%%%%%%%%%%%%%

\section{Numerical results}
\label{Num}

We carried out molecular dynamics simulations to calculate the shear 
viscosity by the Helfand-moment and the Green-Kubo methods. We use 
the standard 6-12 Lennard-Jones 
potential
\begin{equation}
u(r)=4\epsilon\left[\left(\frac{\sigma}{r}\right)^{12}-\left(\frac{\sigma}{r}\right)^{6}\right]
\end{equation}
We 
use the reduced units defined in Table \ref{TABREDUNIT}. 

\begin{table}[h!]
\begin{center}
\begin{tabular}{cc}
\hline
\hline
Quantity	     &	Units \\
\hline
temperature          & $T^* = \frac{k_{\rm B} T}{\epsilon}$ \\
number density       & $n^* = n \sigma^3$ \\
time                 & $t^* = t \sqrt{\frac{\epsilon}{m \sigma^2}}$ \\
distance             & $r^* = \frac{r}{\sigma}$ \\
shear viscosity      & $\eta^* = \eta  \frac{\sigma^2}{\sqrt{m \epsilon}}$\\
\hline
\hline
\end{tabular}
\caption{Reduced units of the Lennard-Jones fluid.}
\label{TABREDUNIT}
\end{center}
\end{table}

All the calculations we perform are done with the cutoff 
$r_c=2.5 \sigma$. The equations of motion are integrated with the 
velocity Verlet algorithm \cite{SABW82} of time step $\Delta 
t=0.003$. The initial positions of the atoms form a fcc lattice 
and the initial velocities are given by a 
Maxwell-Boltzmann 
distribution. Thereafter, the system is equilibrated over $3\times 
10^5$ time steps to reach thermodynamic equilibrium. After the 
equilibration stage, the production stage starts.  At each time step, 
the microscopic flux (\ref{microflux}) and the Helfand moment 
(\ref{helfand-torus}) are calculated. 
Every 300 time units ($10^5$ 
time steps), we compute the time autocorrelation function of the flux 
and the mean-square displacement of the Helfand moment for this piece 
of trajectory and average them with the previous results. Thanks to 
this method we can calculate with a very large statistics since we do 
not need to keep in memory the whole trajectory. Depending on the 
size of the systems ($N=108$-$1372$), the number of pieces of 
trajectory varies between 2000 and 6000, hence the total number of 
time steps is between $2\times 10^8$ and $6\times 10^8$. Statistical 
error is obtained from the mean-square deviation of the correlation 
function or of the mean-square displacement on the trajectory pieces.

\begin{figure}[h!]
\includegraphics[scale=0.6]{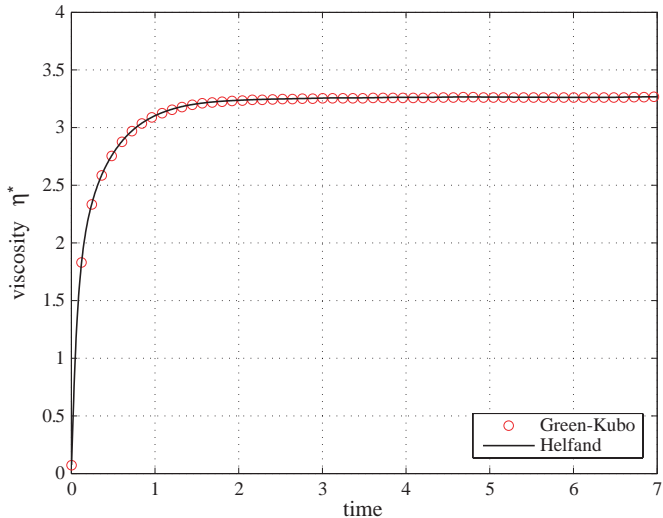}
\caption{Viscosity at the phase point $T^*=0.722$ and $n^*=0.8442$ 
for $N=1372$. The plain line is the derivative of the mean-square 
displacement
of the Helfand moment and the circles the integral of the microscopic 
flux autocorrelation function.}
\label{etacomp}
\end{figure}

We depict in Fig. \ref{etacomp} the time derivative of the 
mean-square displacement of the Helfand moment (\ref{helfand-torus}) 
and the time integral of the autocorrelation function of the 
microscopic flux (\ref{microflux}). In Fig. \ref{etacomp}, the 
calculation is performed for $N=1372$ atoms near the triple point at 
the reduced temperature $T^*=0.722$ and density $n^*=0.8442$. As we 
see, the two method are in perfect agreement.

\begin{figure}[h!]
\includegraphics[scale=0.6]{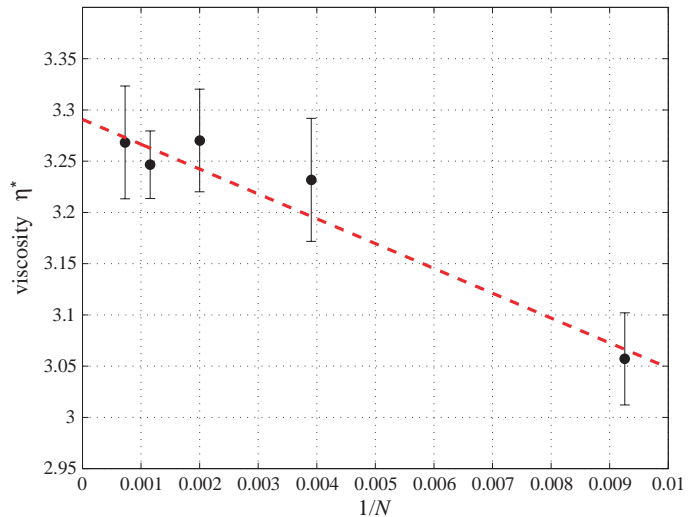}
\caption{Viscosity at the phase point $T^*=0.722$ and $n^*=0.8442$ as 
a function of the inverse of the number $N$ of atoms. The circles are 
the results of the numerical simulations and the dashed line the 
linear extrapolation.}
\label{etaconv}
\end{figure}

We estimated the shear viscosity by a linear fit on the mean-square 
displacement of the Helfand moment. The fit is done in the region 
between 5 and 10 time units to guarantee that the linear regime is 
reached. We depict in Fig. \ref{etaconv} the shear viscosity versus 
the inverse $N^{-1}$ of the system size. The linear extrapolation 
gives the following estimate of shear viscosity
for an infinite system,
\begin{equation}
\eta^*=3.291 \pm 0.057
\label{value}
\end{equation}
This result is in agreement with the previous works as reported in 
Table \ref{TABCOMP}. Indeed, the previous extrapolation found by 
Erpenbeck \cite{erpenbeck-88} $\eta^*=3.345 \pm 0.068$, Palmer 
\cite{palmer94} $\eta^*=3.25 \pm 0.08$, and Meier {\it et al.} 
\cite{MLK-04} $\eta^*=3.258 \pm 0.033$ are all in agreement with our 
result (\ref{value}) within the statistical error.

{\linespread{0.9}
\begin{table}[n]
\begin{center}
\begin{tabular}{lccccccc}
\hline
\hline
Authors     & year  & method   & $r_{\rm cut}^*$   & $T^*$  & $N$ 
& $\eta^*$   & $\Delta \eta^*$\\
\hline
Levesque \textit{et al.} \cite{LVK-73}    & 1973  & GK (MD) & N. C. & 
0.722  & 864      & 4.03     & 0.3   \\
&  & & &       	&   &   &   \\
Ashurst and Hoover 
\cite{ashurst-hoover73}   & 1973  & SSW   & N. C. & 0.722  & 864  & 
3.88  & 0.0026\\
&  & & &       	&   &   &   \\
Ashurst and Hoover 
\cite{ashurst-hoover75} & 1975  & SSW  & N. C. & 0.722  & $\infty$  & 
2.9   & 0.1   \\
&  & & &       	&   &   &   \\
Hoover \textit{et al.} \cite{HEHLAM80} 
& 1980  & SSW  & 2.5 & 0.715  & 864   & 3.0            & 0.15  \\
					       &       & OSW  &   & 
0.722  & 108      & 3.18                & 0.1   \\
&  & & &       	&   &   &   \\
Levesque $^{a)}$ 
& 1980  & GK (MD) & N. C. & 0.728  & 108            & 2.97 
& N. C.  \\
                                                &    & &   & 0.715  & 
256	   	        & 2.92                & N. C.  \\
                                                &   & &    & 0.722  & 
864     	   & 3.85                & N. C.  \\
&  & & &       	&   &   &   \\
Pollock $^{a)}$ 
& 1980  & GK (MD) & N. C. & 0.722  & 256    	      & 2.6 
& 0.1   \\
					       &    &   & & 0.722  & 
500     & 3.2                 & 0.2   \\
&  & & &       	&   &   &   \\
Evans \cite{evans81}                & 
1981  & LE  & 2.5  & 0.722  & 108-256 $^{b)}$    & 3.17            & 
0.03  \\
&  & & &       	&   &   &   \\
Heyes \cite{heyes83} 
& 1983  & DT  & 2.5 & 0.73   & 500                  & 3.08 
& 0.24  \\
&  & & &       	&   &   &   \\
Schoen and Hoheisel 
\cite{schoen-hoheisel85}   & 1985  & GK (MD) & 2.5 & 0.73 & 500 & 
3.18 & 0.15  \\
&  & & &       	&   &   &   \\
Erpenbeck \cite{erpenbeck-88}  & 1988 
& GK (MC)  & 2.5 & 0.722  & 108  & 2.912               & 0.071 \\
					       & &   &    &   & 864 
& 3.200               & 0.160 \\
					       & &   &    &   & 
$\infty$   	& 3.345               & 0.068 \\
&  & & &       	&   &   &   \\
Heyes \cite{heyes-88} 	  & 1988 & GK 
(MD) & N. C. & 0.72   & 108      & 3.2	          & 0.16  \\
					&       &    &   &    & 256 
& 3.5                 & 0.18  \\
					 &      &    &   &    & 500 
& 3.4                 & 0.17  \\
&  & & &       	&   &   &   \\
Ferrario \textit{et al.} \cite{FCHR91} 
	& 1991 & GK (MD) & 2.5 & 0.725  & 500     & 3.02     & 0.07 
\\
                                       &         &    &   & 0.7247 & 
864     	       & 3.24                & 0.10  \\
                                        &        &    &   & 0.725  & 
864     	  & 3.11 - 3.27 $^{c)}$ & 0.10  \\
                                        &        &    &   &   & 2048 
	 & 3.24                & 0.04  \\
                                        &        &    &   &    & 4000 
	  & 3.28                & 0.13  \\
&  & & &       	&   &   &   \\
Palmer \cite{palmer94}	& 1994  & 
TCAF & 2.5 & 0.722  & $\infty$	     & 3.25     	          & 
0.08  \\
&  & & &       	&   &   &   \\
Stassen and Steele 
\cite{stassen-steele95}  & 1995  & GK (MD) & 3.4 & 0.722  & 256 & 
3.297	& N. C.  \\
&  & & &       	&   &   &   \\
Meier \textit{et al.} \cite{MLK-04} 
& 2004  & GEF & 2.5    & 0.722  & 108   & 2.984               & 0.089 
\\
                                                &   &  & 3.25  &   & 
256    	  & 3.105               & 0.093 \\
                                                &    &  & 4.0 &    & 
500    	 & 3.188               & 0.096 \\
                                                &    &  & 5.0 &    & 
864   	  & 3.314               & 0.099 \\
                                                &   &   & 2.5 &    & 
1372       	  & 3.277               & 0.098 \\
                                                &   &   & 5.5 &    & 
2048      	 & 3.224               & 0.097 \\
                                                &    &  & 5.5 &    & 
4000    	& 3.275               & 0.098 \\
                                                &    &   & &    & 
$\infty$  	  & 3.258               & 0.033 \\
&  & & &       	&   &   &   \\
This work    & 2007  & HM (MD)   & 2.5 
& 0.722  & 108   & 3.057 & 0.045 \\
                       &           &             &         & 
& 256   & 3.232 & 0.060 \\
                       &           &              &        & 
& 500   & 3.270 & 0.050 \\
                       &           &             &         & 
& 864   & 3.247 & 0.033 \\
                       &           &              &        & 
& 1372   & 3.268 & 0.055 \\
                       &           &              &        & 
& $\infty$   & 3.291 & 0.029 \\
\hline
\hline
\end{tabular}
\caption{Results found in the literature for the shear viscosity in 
the Lennard-Jones fluid near the triple point. The reduced density 
equals $n^* = 0.8442$, except for the result reported by Heyes (1988) 
\cite{heyes-88} ($n^* = 0.848$), and by Stassen and Steele 
\cite{stassen-steele95} ($n^* = 0.8445$) . Abbreviations: DT, 
difference in trajectories method. GEF, generalized Einstein relation 
with integration of the flux. GK (MC), Green-Kubo results with 
Monte-Carlo method. GK (MD), Green-Kubo results with molecular 
dynamics. OSW, oscillatory shearing walls. SSW, steady shearing 
walls. TCAF, method based on the transverse-current autocorrelation 
functions. HM (MD) the present Helfand-moment method with molecular 
dynamics (MD). The infinite sign means that the value of the shear 
viscosity is obtained by extrapolation for $N \to \infty$. N. C. 
means that the
value has not been communicated.\\
$^{a)}$ Values unpublished but communicated by Hoover \textit{et al.} 
\cite{HEHLAM80}. \\
$^{b)}$ No size dependence. \\
$^{c)}$ Values obtained for different thermostatting rates.}
\label{TABCOMP}
\end{center}
\end{table}
}
%%%%%%%%%%%%%%%%%%%%%%%%%%%%%%%%%%%%%%%%%%%%%%%%%%%%%%%%%%%%%%%%%%%%%%%%%%%%%%%%

\section{Conclusions}
\label{Conclusions}

In this paper, we propose a new method for the computation of shear 
viscosity by molecular dynamics. The \textit{Helfand-moment method} 
is an adaptation of the Helfand formula 
(\ref{Einstein.shear.viscosity}) for systems with periodic boundary 
conditions by adding two terms (\ref{added-terms}) to the original 
expression of the Helfand moment 
(\ref{helfand-moment-shear-viscosity}). The method consists in the 
calculation of the mean-square displacement of the Helfand moment 
(\ref{helfand-torus}).  The variance of this quantity gives the shear 
viscosity by the generalized Einstein relation 
(\ref{Einstein.shear.viscosity}). We have discussed its validity in 
the light of the discussions found in the literature of the beginning 
of the nineties. Thanks to this new method, we have computed the 
shear viscosity in the Lennard-Jones fluid near the triple point. We 
showed that the Helfand-moment method gives the same results as the 
standard Green-Kubo method. Moreover, our extrapolated value of shear 
viscosity is in statistical agreement with those found in the 
literature. More than stating as an alternative method to the 
standard Green-Kubo in equilibrium molecular dynamics, the 
Helfand-moment method is useful and plays a central role in the 
\textit{escape-rate formalism} and the \textit{hydrodynamic-mode 
method}. Indeed,
in these theories, the Helfand moment allows us to put in evidence 
fractal structures at the microscopic level, which are related to the 
transport processes 
\cite{dorf-gasp,gasp-dorf,gasp-book,dorf-book,viscardy-gaspard2}.

We remark that the method can be similarly extended to the bulk 
viscosity $\zeta$. As for the shear viscosity, two terms must be added
to the original expression of the Helfand moment associated with the 
bulk viscosity. Formally, this last coefficient is expressed as 
follows:
\begin{equation}
\zeta + \frac{4}{3} \eta =\lim_{N,V,t \to \infty}\frac{1}{2k_{\rm B} 
T Vt} \left \langle \left[ G^{(\zeta)}(t)-  \langle 
G^{(\zeta)}(t)\rangle\right] ^2\right\rangle ,
\label{Einstein.bulk.viscosity}
\end{equation}
where its Helfand moment is defined as:
\begin{eqnarray}
G^{(\zeta)}(t) &=& \sum_{a} p_{ax}(t) \; x_{a}(t)  \nonumber \\
&-& \sum_{a} \sum_s p_{ax}^{(s)} \; \Delta x_{a}^{(s)} \; 
\theta(t-t_s)  \nonumber \\
&-& \frac{1}{2} \sum_{a \ne b}^{} \int_{0}^{t} d \tau \; F_x ({\bf 
r}_{ab}) \; L_{b|a x} \, ,
\end{eqnarray}
and with $G^{(\zeta)}(0)=0$.
In the companion paper, we will present a similar method for the 
calculation of thermal conductivity \cite{VSG06}.

\acknowledgments
We thank K. Meier for useful discussions.
This research is 
financially supported by the ``Communaut\'e fran\c
caise de Belgique'' (contract ``Actions de Recherche Concert\'ees''
No.~04/09-312) and the National Fund for Scientific Research
(F.~N.~R.~S. Belgium, contract F.~R.~F.~C. No.~2.4577.04).

\appendix

\section{Derivation of the Helfand moment for the shear viscosity in 
periodic systems}
\label{appendixA}

By taking the time derivative of the Helfand moment (\ref{Helfand+I}), we have:
\begin{eqnarray}
\frac{d G^{(\eta)}(t)}{dt} &=& \sum_{a}^{} 
\frac{p_{ax}(t)p_{ay}(t)}{m} \nonumber \\
&+& \sum_{a}^{} \sum_{s}^{} p_{ax}(t) \Delta y_{a}^{(s)} \delta 
(t-t_s) \nonumber \\
&+& \sum_{a\ne b}^{} F_{x}(\mathbf{r}_{ab}) y_{a}(t)
+ \frac{dI(t)}{dt}
\label{current-I}
\end{eqnarray}
where we have used the modified Newton equations (\ref{Newton}). The 
term implying the interparticle force $\mathbf{F}({\bf r}_{ab})$ may 
be modified into
\begin{equation}
\sum_{a\ne b}^{} F_{x}(\mathbf{r}_{ab}) y_{a}(t) = 
\frac{1}{2}\sum_{a\ne b}^{} F_x({\bf r}_{ab}) y_{a}
+ \frac{1}{2}\sum_{a\ne b}^{} F_x({\bf r}_{ba}) y_{b} \, .
\end{equation}
Since the force $\mathbf{F}$ is central, we obtain $F_{x}({\bf 
r}_{ab}) = - F_x({\bf r}_{ba})$, which
implies that
\begin{equation}
\sum_{a\ne b}^{} F_{x}(\mathbf{r}_{ab}) y_{a}(t) = 
\frac{1}{2}\sum_{a\ne b}^{} F_x({\bf r}_{ab}) \, (y_{a}-y_{b})  \, .
\end{equation}
which still differs from the corresponding term 
appearing in the microscopic flux (\ref{microflux}) defined with the 
minimum-image convention because $y_{a}-y_{b}=y_{ab}+L_{b|a y}$ 
according to Eqs. (\ref{MIC}) and (\ref{definition-Lab}).
Consequently, Eq. (\ref{current-I}) becomes
\begin{eqnarray}
\frac{d G^{(\eta)}(t)}{dt} &=& J^{(\eta)} (t) + \frac{1}{2} \sum_{a,b 
\ne a}^{} F_x ({\bf r}_{ab}) L_{b|a y} \nonumber \\
&+& \sum_{a}^{} \sum_{s}^{} p_{ax}(t) \Delta y_{a}^{(s)} \delta (t-t_s)
+ \frac{dI(t)}{dt} \, . \nonumber \\
\end{eqnarray}
Comparing with Eq. (\ref{derivativeG}), we should have
\begin{eqnarray}
\frac{dI(t)}{dt}  &=& - \sum_{a}^{} \sum_{s}^{} p_{ax}(t) \Delta 
y_{a}^{(s)} \delta (t-t_s) \nonumber \\
&-& \frac{1}{2} \sum_{a,b \ne a}^{} F_x ({\bf r}_{ab}) L_{b|a y}
\end{eqnarray}
whereupon $I(t)$ can be expressed as
\begin{eqnarray}
I(t) &=& - \sum_{a}^{} \sum_{s}^{} p_{ax}^{(s)} \Delta y_{a}^{(s)} 
\theta (t-t_s) \nonumber \\
&-& \frac{1}{2} \sum_{a,b \ne a}^{} \int_{0}^{t} d \tau \; F_x ({\bf 
r}_{ab}) \; L_{b|a y} \; .
\end{eqnarray}

%%%%%%%%%%%%%%%%%%%%%%%%%%%%%%%%%%%%%%%%%%%%%%%%%%%%%%%%%%%%%%%%%%%%%%%%%%%%%%%%%%%%%%%%%%%%%%%%%%%%%%%%%%%%%%%%%%%%%%%%%%%%%%%%%%%%%

%\bibliography{bibliobibtex}

\end{document}